\documentclass{cjaa}                   

\usepackage{graphicx}                   
\input{epsf.sty}                        
\input{psfig.sty}                       

\newcommand{\ls}{LS~5039}
\newcommand{\egls}{3EG~J1824$-$1514}
\newcommand{\lsi}{LS~I~+61~303}
\newcommand{\cg}{2CG~135+01}
\newcommand{\eglsi}{3EG~J0241$+$6103}

\begin{document}

   \title{High-energy $\gamma$-ray Emission from Microquasars:}
   \subtitle{\ls\ and \lsi}

   \volnopage{Vol.0 (200x) No.0, 000--000}      
   \setcounter{page}{1}          

   \author{V. Bosch-Ramon\mailto{vbosch@am.ub.es}
   \and J. M. Paredes
      }
   \offprints{V. Bosch-Ramon}                   

   \institute{Departament d'Astronomia i Meteorologia, Universitat de Barcelona, 
              Av. Diagonal 647, E-08028 Barcelona, Spain\\
             \email{vbosch@am.ub.es}
          }

   \date{?; ?}

   \abstract{      
The possible associations between the microquasars \ls\ and \lsi\ and the EGRET sources \egls\
and \eglsi\ suggest that microquasars could also be sources of high-energy  $\gamma$-rays. In
this work, we present a detailed numerical inverse Compton (IC) model, based on a  microquasar
scenario, that reproduces the high-energy $\gamma$-ray spectra and variability observed by EGRET
for the mentioned sources. Our model considers a population of relativistic electrons
entrained in a cylindrical inhomogeneous jet that interact through IC scattering with both the
radiation and the magnetic fields. 
   \keywords{X-rays: binaries --- Stars: LS~5039, LS~I~+61~303 --- gamma-rays: observations 
   --- gamma-rays: theory  }
   }

   \authorrunning{V. Bosch-Ramon \& J. M. Paredes}           
   \titlerunning{High-energy $\gamma$-ray Emission in Microquasars}  

   \maketitle

\section{Introduction}           
\label{sect:intro}

Microquasars are a selected class of X-ray binaries that produce relativistic radio jets
(Mirabel \& Rodr{\'{\i}}guez \cite{Mirabel&rodriguez99}, Fender \cite{Fender04}). The origin
of the jets is related to the matter accreted by the compact object, a neutron star or a
black hole, from the companion star. These systems behave as scaled-down versions of quasars
and active galactic nuclei. The population of microquasars is still a very reduced one, with
about sixteen known objects up to now (Rib\'o \cite{Ribo02t}). Due to the presence in these
systems of both a relativistic jet with a population of highly relativistic leptons and
strong radiation fields, microquasars are promising $\gamma$-ray emitter candidates through IC
interaction (for synchrotron self Compton (SSC) models, see, i.e. Atoyan \& Aharonian
\cite{Atoyan&aharonian99}; for external Compton (EC) models, see, i.e. Kaufman Bernad\'o
et~al. \cite{Kaufman-Bernado02}), although these sources could emit at such energy ranges by
other means (i.e. see the hadronic model of Romero et~al. \cite{Romero03}). We are interested
on determining whether a microquasar jet emitting by IC scattering can reproduce the EGRET
data of the two sources associated to microquasars: \egls/\ls\ and \eglsi/\lsi. Our numerical
model considers a population of relativistic electrons entrained in a cylindrical
inhomogeneous jet interacting with both the radiation and the magnetic fields, accounting for
the synchrotron, the EC and the SSC electron energy losses. 

\section{\ls\ and \lsi}
\label{sect:}

\ls\ is a High Mass X-ray binary system (HMXB) located at 2.9~kpc from Earth (Rib\'o et~al.
\cite{Ribo02}). The stellar companion is a bright (V~$\sim$~11) star of ON6.5~V((f)) spectral
type, the compact object seems to be a neutron star, the orbital period of the system is
$P=4.4267 \pm 0.0005$ days, the eccentricity, $e=0.48 \pm 0.06$, and the semi-major axis of
the orbit $a=2.6\times10^{12}$~cm (McSwain et~al. \cite{Mcswain04}). The microquasar nature
of \ls\ was clearly established when non-thermal radiation produced in a mildly relativistic
jet was detected, being also proposed as the counterpart of the unidentified $\gamma$-ray
source \egls\ (Paredes et~al. \cite{Paredes00}). 

\lsi\ is another HMXB whose optical counterpart is a bright (V$\sim$10.8) star of B0 V
spectral type (Paredes \& Figueras \cite{Paredes&figueras86}), presenting and equatorial disk
(Be star) and harbouring likely a neutron star as a compact object (Hutchings \& Crampton
\cite{Hutchings&crampton81}). This source is at a distance of about 2~kpc. The more accurate
value for the orbital period is $P=26.4960 \pm 0.0028$~days (Gregory \cite{Gregory02}), the
eccentricity is $e\sim 0.7$ (Casares et~al. \cite{Casares04}), and the orbital semi-major
axis is about $a=5\times10^{12}$~cm. This source presents variability at different
timescales, from radio to $\gamma$-rays (see, i.e., Taylor \& Gregory \cite{Taylor&gregory82},
Paredes \& Figueras \cite{Paredes&figueras86}, Goldoni \& Mereghetti
\cite{Goldoni&mereghetti95}, Tavani et~al. \cite{Tavani98}, Massi \cite{Massi04}). The
microquasar nature of \lsi\ was established when a mildly relativistic jet structure was
detected through VLBI observations of this source (Massi et~al. \cite{Massi01},
\cite{Massi04b}), althought this object had already been proposed to be the counterpart of
the $\gamma$-ray source \cg/\eglsi\ (Gregory \& Taylor \cite{Gregory&taylor78}, Kniffen et~al.
\cite{Kniffen97}). 

\section{Modelling the $\gamma$-ray Emission from Microquasars}
\label{sect:Mod}

In this model, we assume that the leptons of the jet dominate the radiative processes related
to the $\gamma$-ray production. The relativistic population of electrons, already accelerated
and flowing away into the jet, is exposed to external photons as well as to the synchrotron
photons emitted by the electrons, since we take into account the magnetic field in our model.
The $\gamma$-ray emitting region, the $\gamma$-jet, is assumed to be closer to the compact
object than the observed radio jets. This $\gamma$-jet is supposed to be short enough to be
considered cylindrical. The magnetic field ($B_{\gamma}$) has been taken  to be constant, as
an average along the jet. The energy losses of the relativistic leptonic plasma at the
$\gamma$-jet are mainly due to synchrotron emission, SSC scattering, and EC scattering. Due to
the importance of the losses, the electron energy distribution density along the $\gamma$-jet
model varies significantly, thus the $\gamma$-jet is studied by splitting it into cylindrical
transverse cuts or slices. 

Regarding IC interaction, we have used the cross-section of Blumenthal \& Gould
(\cite{Blumenthal&gould70}), $\sigma(x,\epsilon_0,\gamma_{\rm e})$, which takes into account
the Thomson and the Klein-Nishina regimes of interaction; $\epsilon_0$ is the seed  photon
energy, $\gamma_{\rm e}$ is the scattering electron Lorentz factor, and $x$ is actually  a
function which depends on both of the former quantities and on the scattered photon energy
($\epsilon$). The electron distribution is assumed to be initially a power law
($N(\gamma_{\rm e})\propto E^{-p}$, where $E$ is the electron energy), which evolves under
the conditions imposed by the magnetic and the radiation fields. Thus, the electron
distribution function of a certain slice ($N(\gamma_{\rm e},z)$) depends on both the distance
to the compact object ($z$) and $\gamma_{\rm e}$. The components of the total seed photon
radiation field ($U(\epsilon_0,z)$) are any present external radiation field ($U_{\rm
ext}(\epsilon_0,z)$) and the synchrotron radiation field produced by the jet's relativistic
electrons, all  of them in the jet's reference frame (for the external photon fields, see
Dermer  \& Schlickeiser \cite{Dermer&schlickeiser02}). It must be noted that, since \ls\ and
\lsi\ are not strong X-ray emitters, we will assume that the most important source of seed
photons is the companion star, neglecting at the present stage the disk and corona 
contributions (see Bosch-Ramon et al. \cite{bosch04}). 
The formulae that represent $U(\epsilon_0,z)$ and its components (a black-body stellar 
photon field and a synchrotron photon field) are well 
described in the work of Bosch-Ramon \& Paredes \cite{Bosch-Ramon&paredes04}.

The free parameters of the model are $B_{\gamma}$ and the maximum electron Lorentz factor at
the slice closest to the compact object ($\gamma_{\rm e0}^{\rm max}$). The leptonic kinetic
luminosity or leptonic jet power ($L_{\rm ke}$) is set free also, and it is scaled with the
observed luminosity, in order to reproduce the observations. The jet radius, although it is
not well constrained, has been fixed to several electron Larmor radii ($\sim 10^7$~cm), and
the electron power-law index of the injected distribution has been deduced from radio
observations (for \ls, see Mart{\'{\i}} et~al. \cite{Marti98}; for \lsi, see Ray et~al.
\cite{Ray97}). The luminosity per energy unit in the reference frame of the jet
($L_{\epsilon}$) is presented in  Eq.~\ref{eq:lumicSRj}. The photon flux per energy unit or
spectral photon distribution in the reference frame of the observer ($I'_{\epsilon'}$) is
shown in  Eq.~\ref{eq:lumicSRobs}. The magnitudes with ($'$) are in the observer reference
frame.
\begin{eqnarray}
L_{\epsilon}&=&\epsilon\sum^{z_{\rm max}}_{z_{\rm min}} V_{\rm slice}(z)
\int^{\epsilon_0^{\rm max}(z)}_{\epsilon_0^{\rm min}(z)}
\int^{\gamma_{\rm e}^{\rm max}(z)}_{\gamma_{\rm e}^{\rm min}(z)}
\frac{U(\epsilon_0,z)}{\epsilon_0}
N(\gamma_{\rm e},z)
\frac{d\sigma(x,\epsilon_0,\gamma_{\rm e})}{d\epsilon}
d\gamma
d\epsilon_0
\label{eq:lumicSRj}
\end{eqnarray}
\begin{equation}
I'_{\epsilon'}=\frac{\delta^{2+p}}{4\pi D^2 \epsilon'}L_{\epsilon'}
\label{eq:lumicSRobs}
\end{equation}
where $\delta$ is the Doppler factor of the jet, $D$ is the distance from the microquasar to
the observer, and $V_{\rm slice}(z)$ is the volume of the slice at a distance $z$ 
from the compact object. For further details of the model, see Bosch-Ramon \& Paredes
(\cite{Bosch-Ramon&paredes04}, \cite{Bosch-Ramon&paredes04b}).

\section{Results and conclusions}
\label{sect:conc}

\begin{figure}
   \vspace{2mm}
   \begin{center}
   \hspace{3mm}\psfig{figure=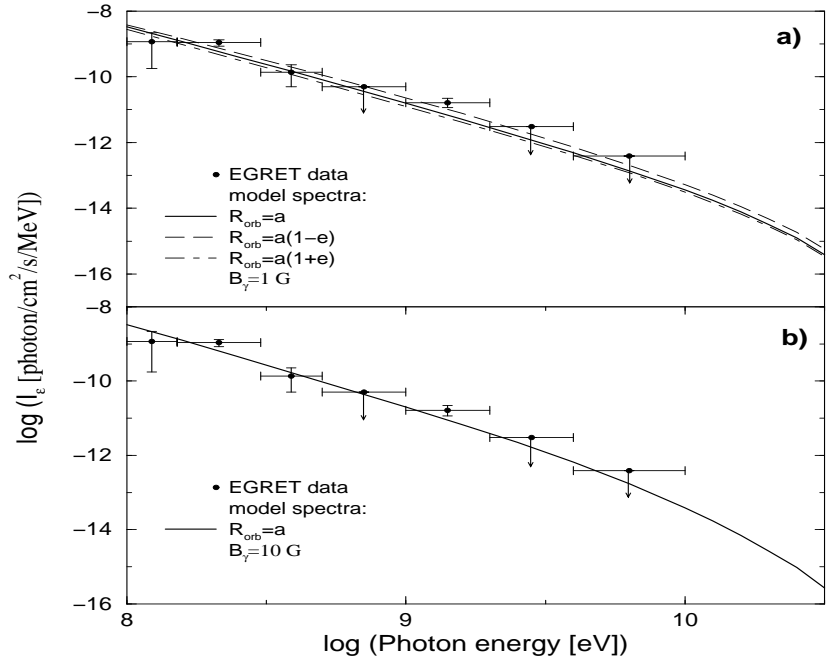,width=110mm,height=90mm,angle=0.0}

   \parbox{180mm}{{\vspace{2mm} }}

   \caption{
Computed IC spectral photon
distribution above 100~MeV for \ls\ using the physical parameter values presented in 
Table~\ref{Tab:par}. The EGRET data points are also shown. The upper limits on
undetected EGRET points are plotted with arrows. Only stellar radiation density changes 
due to the orbital eccentricity have been taken into account.
 \textbf{a)} A magnetic field of 1~G has
been adopted. Also, the IC spectral photon distribution for both the apastron and the periastron
passage are shown. \textbf{b)} A magnetic field of 10~G has been adopted.   
   }
   \label{Fig:ls}
   \end{center}
\end{figure}

\begin{figure}
   \vspace{2mm}
   \begin{center}
   \hspace{3mm}\psfig{figure=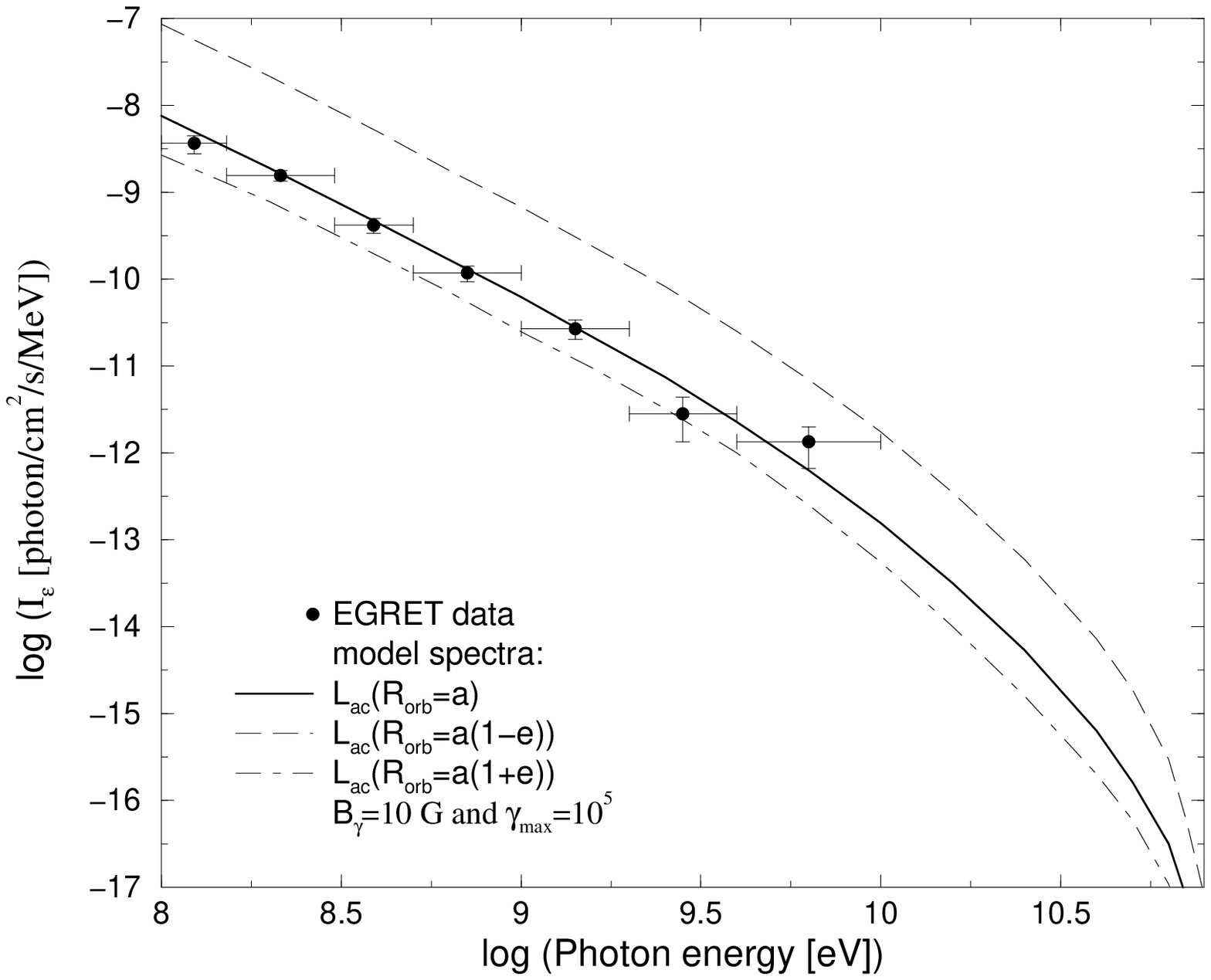,width=110mm,height=80mm,angle=0.0}

   \parbox{180mm}{{\vspace{2mm} }}

   \caption{
Computed spectral photon 
distribution above 100~MeV for \lsi\ plotted with the EGRET data points. We have 
used the physical parameter values presented in Table~\ref{Tab:par}. 
Unlike in Fig.~\ref{Fig:ls}, the spectral shape and the 
variability are the same for both $B_{\gamma}$=1 and 10~G. Thus there is only one plot.
There are plotted the
computed $I'_{\epsilon'}$ for different orbital radii: $a$ (solid line), the distance
at the periastron passage ($a(1-e)$, dashed line), and the distance at the apastron
passage ($a(1+e)$, dotted line). Now, accretion changes due to orbital eccentricity have been
included in calculations.
   }
   \label{Fig:lsi}
   \end{center}
\end{figure}

We have applied our model to both \ls\ and \lsi\ in order to reproduce the spectra observed
by EGRET. In Table~\ref{Tab:par}, the known parameter values as well as the adopted values 
for the free parameters are presented.  The computed spectral photon distributions are shown
in the Figs.~\ref{Fig:ls} and  \ref{Fig:lsi} for \ls\ and \lsi, respectively. The results
show that these microquasars can produce high-energy $\gamma$-rays through IC interaction
between a relativistic leptonic population, entrained in an inhomogeneous cylindrical jet,
and the external and synchrotron photon fields. 

Our model can reproduce EGRET data with reasonable parameter values, and variability is
naturally expected from stellar radiation and accretion changes due to orbital eccentricity 
(precession can introduce also variability, see Massi et~al. \cite{Massi04b}). Since, at the
present stage, no clear varying accretion influence has been detected in \ls\ at lower
energies, we have not introduced it to compute the presented spectral photon distributions
(see Fig.~\ref{Fig:ls}), but only the eccentric orbit effects on the stellar radiation
density. However, since orbital eccentricity is clearly affecting radio (Taylor \& Gregory
\cite{Taylor&gregory82}) and X-ray emission (Goldoni \& Mereghetti
\cite{Goldoni&mereghetti95}) in the case of \lsi, we have taken into account it to calculate
its emission at $\gamma$-rays. The magnetic field can control the loss of energy of the
electrons by SSC losses. Therefore, with two magnetic field strenghts: 1 and 10 G, the
emission turns from being dominated by EC effect to be dominated the SSC effect. It is shown
in Fig.~\ref{Fig:ls}, where emission is not affected by the eccentricity of the orbit when
the magnetic field is high, since changes in the stellar photon density do not affect the
spectrum in the SSC dominant regime  (accretion changes are not included in the
calculations). It is worth noting also that in the case of \lsi, where accretion changes are
taken into account, they are far more important than the stellar photon density variations,
implying almost the same effects on the spectrum variability for both cases, the EC and the
SSC ones.

These results strengthen the idea that microquasars are $\gamma$-ray emitters, being \ls\ and
\lsi\ likely counterparts of the two EGRET sources \egls\ and \eglsi\ respectively. However, further
observations with the new generation of $\gamma$-ray instruments (i.e. AGILE, GLAST, HESS,
MAGIC, etc...) are mandatory in order to get enough angular resolution, flux and temporal
sensitivity and energy range width to determine properly the association
microquasars/EGRET sources and to better constrain the current models of emission.

\begin{table}[]
  \caption[]{Parameter values for \ls\ and \lsi}
  \label{Tab:par}
  \begin{center}\begin{tabular}{clcl}
  \hline\noalign{\smallskip}
Parameter & Adopted values for \\
 & \ls\ / \lsi \\
  \hline\noalign{\smallskip}
Jet Lorentz factor ($\Gamma_{\rm jet}$) & 1.02 / 1.25 \\ 
Angle between the jet and the observer line of sight ($\theta$) & $30^{\circ}$ \\
Jet velocity ($v_{\rm jet}$) & 0.2$c$ / 0.6$c$ \\
Orbital semi-major axis ($a$) & $2.6\times10^{12}$ / $5\times10^{12}$~cm \\
Orbital eccentricity ($e$) & 0.5 / 0.7 \\
Distance to the observer ($D$) & 2.9 / 2 kpc \\
Star total luminosity ($L_{\rm star}$) & $1.2\times10^{39}$ / $2\times10^{38}$~erg~s$^{-1}$ \\
Photon flux at the EGRET band ($I_{\rm >100~MeV}$) & $3.3\times10^{-7}$ / $8\times10^{-7}$~photon~cm$^{-2}$~s$^{-1}$ \\
Photon index at the EGRET band ($\Gamma$) & $\sim$2.2 \\
Jet radius ($R_{\gamma}$) & $10^7$~cm \\
Injected electron power-law index ($p$) & 2 / 1.7 \\
  \hline\noalign{\smallskip}
Free parameter & Adopted values for \\ 
  & \ls\ / \lsi \\
  \hline\noalign{\smallskip}  
Magnetic field ($B_{\gamma}$) & 1--10~G\\
Leptonic jet power ($L_{\rm ke}$) & $10^{35}$--$3\times10^{35}$ / $10^{36}$--$3\times10^{36}$~erg~s$^{-1}$ \\
Maximum electron Lorentz factor ($\gamma_{\rm e0}^{\rm max}$) & $10^5$ \\
  \noalign{\smallskip}\hline
  \end{tabular}\end{center}
\end{table}



\begin{acknowledgements}

V.B-R. and J.M.P. acknowledge partial support by DGI of the Ministerio de Ciencia y
Tecnolog{\'{\i}}a (Spain) under grant AYA-2001-3092, as well as additional support from the
European Regional Development Fund (ERDF/FEDER). During this work, V.B-R has been supported
by the DGI of the Ministerio de Ciencia y Tecnolog{\'{\i}}a (Spain) under the fellowship
FP-2001-2699. 

\end{acknowledgements}

\label{lastpage}

\end{document}